\documentclass[12pt]{article}

\usepackage{epsfig,amsmath,amssymb,verbatim,mathrsfs}
\usepackage[square, comma, sort&compress]{natbib}

\def\beq{\begin{eqnarray}}
\def\eeq{\end{eqnarray}}
\def\bea{\begin{eqnarray}}
\def\eea{\end{eqnarray}}

\def\tev{\, {\rm TeV}}
\def\gev{\, {\rm GeV}}

\newcommand{\gsim}{\lower.7ex\hbox{$\;\stackrel{\textstyle>}{\sim}\;$}}
\newcommand{\lsim}{\lower.7ex\hbox{$\;\stackrel{\textstyle<}{\sim}\;$}}

\def\stilde{\widetilde}

\newcommand{\newc}{\newcommand}
\newc{\Nc}{N_{c}}
\newc{\CG}{C_G}
\newc{\gp}{g'}
\newc{\stopi}{\stilde t_i}
\newc{\sboti}{\stilde b_i}
\newc{\staui}{\stilde \tau_i}
\newc{\stopj}{\stilde t_j}
\newc{\sbotj}{\stilde b_j}
\newc{\stauj}{\stilde \tau_j}
\newc{\stopI}{\stilde t_1}
\newc{\stopII}{\stilde t_2}
\newc{\sbotI}{\stilde b_1}
\newc{\sbotII}{\stilde b_2}
\newc{\stauI}{\stilde \tau_1}
\newc{\stauII}{\stilde \tau_2}
\newc{\sstop}{s_{t}}
\newc{\cstop}{c_{t}}
\newc{\ssbot}{s_{b}}
\newc{\csbot}{c_{b}}
\newc{\sstau}{s_{\tau}}
\newc{\cstau}{c_{\tau}}
\newc{\Sstop}{s_{2t}}
\newc{\Cstop}{c_{2t}}
\newc{\Ssbot}{s_{2b}}
\newc{\Csbot}{c_{2b}}
\newc{\Sstau}{s_{2\tau}}
\newc{\Cstau}{c_{2\tau}}
\newc{\salpha}{s_\alpha}
\newc{\calpha}{c_\alpha}
\newc{\Calpha}{c_{2\alpha}}
\newc{\Salpha}{s_{2\alpha}}
\newc{\sbetapm}{s_{\beta_\pm}}
\newc{\cbetapm}{c_{\beta_\pm}}
\newc{\Sbetapm}{s_{2 \beta_\pm}}
\newc{\Cbetapm}{c_{2 \beta_\pm}}
\newc{\sbetaO}{s_{\beta_0}}
\newc{\cbetaO}{c_{\beta_0}}
\newc{\SbetaO}{s_{2 \beta_0}}
\newc{\CbetaO}{c_{2 \beta_0}}
\newc{\vu}{v_u}
\newc{\vd}{v_d}
\newc{\seL}{\stilde e_L}
\newc{\smuL}{\stilde \mu_L}
\newc{\seR}{\stilde e_R}
\newc{\smuR}{\stilde \mu_R}
\newc{\suL}{\stilde u_L}
\newc{\sdL}{\stilde d_L}
\newc{\suR}{\stilde u_R}
\newc{\sdR}{\stilde d_R}
\newc{\scL}{\stilde c_L}
\newc{\ssL}{\stilde s_L}
\newc{\scR}{\stilde c_R}
\newc{\ssR}{\stilde s_R}
\newc{\snue}{\stilde \nu_e}
\newc{\snumu}{\stilde \nu_\mu}
\newc{\snutau}{\stilde \nu_\tau}
\newc{\Gpm}{G^\pm}
\newc{\Hpm}{H^\pm}
\newc{\FFbS}{\overline{FF}S}
\newc{\FFbV}{\overline{FF}V}
\newc{\FSS}{F_{SS}}
\newc{\FSSS}{F_{SSS}}
\newc{\FFFS}{F_{FFS}}
\newc{\FFFbS}{F_{\overline{FF}S}}
\newc{\FSSV}{F_{SSV}}
\newc{\FVS}{F_{VS}}
\newc{\FVVS}{F_{VVS}}
\newc{\FFFV}{F_{FFV}}
\newc{\FFFbV}{F_{\overline{FF}V}}
\newc{\Fgauge}{F_{\rm gauge}}
\newc{\DRbarprime}{$\overline{\rm DR}'$ }
\newc{\DRbar}{$\overline{\rm DR}$ }
\newc{\MSbar}{$\overline{\rm MS}$ }
\newc{\Yu}{{\bf Y}_u}
\newc{\Yd}{{\bf Y}_d}
\newc{\Ye}{{\bf Y}_e}
\newc{\Au}{{\bf a}_u}
\newc{\Ad}{{\bf a}_d}
\newc{\Ae}{{\bf a}_e}
\newc{\bm}{{\bf m}}
\newc{\zhol}{Z^{\rm hol}}

\newcommand{\nnmb}{\nonumber}

\begin{document}

\setlength{\baselineskip}{0.2in}



\begin{titlepage}
\noindent
\begin{flushright}
KITP-08-19\\
MCTP-09-04\\
\end{flushright}
\vspace{1cm}

\begin{center}
  \begin{Large}
    \begin{bf}
Higgs Boson Signatures of MSSM Electroweak Baryogenesis\\
     \end{bf}
  \end{Large}
\end{center}
\vspace{0.2cm}

\begin{center}

\begin{large}
Arjun Menon$^{a}$ and David E. Morrissey$^{b}$\\
\end{large}
\vspace{0.3cm}
  \begin{it}
$^a$Michigan Center for Theoretical Physics (MCTP) \\
Physics Department, University of Michigan, Ann Arbor, MI 48109
\vspace{0.5cm}\\
$^b$
Jefferson Physical Laboratory, Harvard University,\\
Cambridge, Massachusetts 02138, USA
\vspace{0.5cm}
\end{it}\\

\end{center}

\center{\today}

\begin{abstract}

  Electroweak baryogenesis (EWBG) in the MSSM can account for the 
cosmological baryon asymmetry, but only within a restricted
region of the parameter space.  In particular, MSSM EWBG requires
a mostly right-handed stop that is lighter than the top quark
and a standard model-like light Higgs boson.  In the present work 
we investigate the effects of the light stop on Higgs boson 
production and decay.  Relative to the standard model Higgs boson, 
we find a large enhancement of the Higgs production rate through
gluon fusion and a suppression of the Higgs branching fraction into
photon pairs.  These modifications in the properties of the Higgs boson 
are directly related to the effect of the light stop on the
electroweak phase transition, and are large enough that they can
potentially be tested at the Tevatron and the LHC.

\end{abstract}

\vspace{1cm}

\end{titlepage}

\setcounter{page}{2}


\vfill\eject



\newpage

\section{Introduction}

  Supersymmetry is a well-motivated solution to the gauge 
hierarchy problem~\cite{Martin:1997ns}.  Within the minimal
supersymmetric extension of the Standard Model~(SM), the MSSM,
it is remarkable that one also finds an excellent unification 
of gauge couplings, a candidate for the dark matter, and the 
potential to generate the baryon asymmetry of the universe.  
In particular, the baryon asymmetry can arise in the MSSM through 
the mechanism of electroweak baryogenesis~(EWBG)~\cite{
Kuzmin:1985mm,Cohen:1993nk}. 

  While MSSM EWBG can account for the baryon excess, 
it can only do so within a specific and tightly constrained 
region of the MSSM parameter space.  Successful EWBG requires 
a significant new source of CP violation beyond the CKM phase 
of the SM~\cite{Gavela:1993ts,Huet:1994jb}, and a strongly 
first-order electroweak phase transition~\cite{Kuzmin:1985mm}.  
Together, these requirements, along with corresponding phenomenological 
bounds, largely fix the superpartner spectrum of the MSSM.

  New sources of CP violation are strongly constrained by 
searches for permanent electric dipole moments~(EDMs) of atoms, 
neutrons, and electrons~\cite{RamseyMusolf:2006vr}.  
MSSM EWBG requires a significant phase 
in a light electroweak gaugino-higgsino (chargino and neutralino) 
sector~\cite{Carena:1997gx,Multamaki:1997ep,
Cline:1997vk,Riotto:1997gu}.\footnote{
Contributions from CP-violating phases in the squark sector
are suppressed by the heavy masses of these states~\cite{Carena:1997gx},
even when these phases are flavour-dependent~\cite{Delepine:2002as}.}  
Such phases induce EDMs at one-loop order through quantum corrections 
involving first- and second-generation squarks 
and sleptons~\cite{Abel:2001vy}. 
To avoid the experimental bounds on permanent EDMs, 
these sfermions must be heavier than about 
$10\,\tev$~\cite{Murayama:2002xk,Balazs:2004ae,Lee:2004we,Cirigliano:2006wh}.  
At two-loop order there arise quantum corrections 
involving charginos and Higgs bosons that cannot be decoupled 
in this way~\cite{Chang:1998uc,Chang:2005ac,Giudice:2005rz,Li:2008kz}, 
further constraining the allowed MSSM parameter 
space~\cite{Pilaftsis:2002fe,
Konstandin:2005cd,Li:2008ez,Balazs:2004ae,Lee:2004we,Cirigliano:2006wh,
Chang:1998uc}.

  Successful EWBG also requires a strongly first-order electroweak phase 
transition.  This can be achieved within the MSSM if the lightest Higgs boson
is SM-like, and one of the scalar top quarks (stops) is 
very light~\cite{Carena:1996wj,Delepine:1996vn}.  Precision 
electroweak constraints force the light stop to be mostly right-handed.  
On the other hand, the LEP-II bound on the light SM-like Higgs boson 
mass of $114.7\,\gev$~\cite{Amsler:2008zzb} implies that the
second mostly left-handed stop must be much heavier.  
A recent re-analysis of the electroweak phase transition 
within the MSSM indicates that the acceptable stop 
and Higgs sector parameters lie within the ranges~\cite{Carena:2008vj}
\beq
\begin{array}{cc}
(-150\,\gev)^2 \lesssim m_{\tilde{t}_R}^2  \lesssim -(50\,\gev)^2,&~~~~~
~~0\lesssim |{A_t - \mu\,\cot\beta}|/m_{Q_3} \lesssim 0.7,\\
~m_{Q_3}^2\gtrsim (6\,\tev)^2,&~~~~~ 
5\lesssim \tan\beta \lesssim 15. 
\end{array}
\eeq
The tachyonic right-handed stop mass implies that the lightest 
stop mass eigenstate is lighter than the top quark.
It also implies that the standard electroweak vacuum is only
metastable against decay to a deeper colour-breaking vacuum,
but this is acceptable provided the electroweak vacuum forms
first and is sufficiently long-lived~\cite{Carena:1997gx,Carena:2008vj}.
In the region consistent with viable EWBG, the light stop mass
as well as the light Higgs boson mass are both found to 
lie below $125\,\gev$~\cite{Carena:2008vj}.

  Together, the dual requirements of new CP-violation and a strongly
first-order electroweak phase transition largely fix the MSSM 
super particle spectrum: all sfermions other than the mostly right-handed stop
(and possibly a right-handed sbottom) must be very heavy, while the
electroweak gauginos and higgsinos must remain relatively light.
This spectrum is challenging to study at the LHC.  The first- and 
second-generation squarks and sleptons are typically too heavy to be 
produced directly or to play a significant role in decay cascades.  
Some information can be obtained about the electroweak gauginos and 
higgsinos from electroweak production~\cite{Cirigliano:2006wh}, 
although the reach is severely diminished if these states 
decay significantly into the light stop~\cite{Carena:2006gb}.
This leaves the light stop, and possibly a lighter
gluino, as the primary sources of LHC signals.  

  Despite being produced very abundantly, 
a light mostly-right-handed stop is difficult to probe 
at the LHC.  Direct searches for a light stop at the Tevatron require 
this state to be relatively degenerate with the lightest supersymmetric
particle~(LSP) or to decay primarily 
into three- or four-body modes~\cite{Demina:1999ty,Balazs:2004bu,
Aaltonen:2007sw,Abazov:2008rc}.  
Indeed, a light stop that is nearly degenerate with a mostly-Bino 
neutralino LSP can lead to an acceptable thermal relic density
through co-annihilation between these states~\cite{Balazs:2004bu,
Balazs:2004ae}.
This near-degeneracy implies that the decay products of the light 
stop will be soft and difficult to tag.  

    A number of studies have investigated ways to probe this 
light stop scenario at the LHC~\cite{Kraml:2005kb,Martin:2008aw,
Bhattacharyya:2008tw,Carena:2008mj,Hiller:2008wp,Martin:2008sv}. 
Ref.~\cite{Kraml:2005kb} proposed 
an indirect search for light stops through di-gluino production with decays
into same-sign top quarks.  With sufficient luminosity, this provides
an LHC discovery mode for gluino masses up to about $1000\,\gev$,
although parameter determination is 
challenging~\cite{Kraml:2005kb,Martin:2008aw}.  
Making use of the very large stop production rate at the LHC,
Ref.~\cite{Carena:2008mj} proposed a search for stops in 
association with a hard photon or a gluon jet.
The additional tag provides a trigger which can be combined with
a cut on missing $E_T$ in the event to provide a viable 
mono-jet or mono-photon signature of the light stop  
when it is nearly degenerate with the LSP~\cite{Carena:2008mj}.
With a very small mass difference between the stop and a neutralino
LSP, the flavour-violating decay mode $\tilde{t}_1\to c\tilde{\chi}_1^0$
can lead to a displaced vertex in the context of minimal flavour 
violation~\cite{Hiller:2008wp}.
Light stops can also form stoponium, which may potentially be detected
through its decays to diphotons~\cite{Martin:2008sv}.

  In the present work we show that the light stop required 
for EWBG in the MSSM leads to significant modifications of
the production and decay rates of the Higgs boson relative 
to the SM.  It is well known that new light coloured 
particles, and the scalar tops of the MSSM in particular, 
can have a significant effect on the Higgs production rate 
through gluon fusion~\cite{Kileng:1995pm,Kane:1995ek,Dawson:1996xz,
Dermisek:2007fi,Low:2009nj}.
The light stop required for electroweak baryogenesis
is unique in that it pushes these effects to their limits.
On account of the large hierarchy between the masses of
the heavy sfermions and the light stop implied by MSSM EWBG,
we make use of an effective theory in which the heavy states
are integrated out explicitly to compute the low-energy
Higgs-stop couplings~\cite{Carena:2008rt}.   
The changes in the properties of the Higgs boson induced
by the light stop are directly related to the effects of
the light stop on the electroweak phase transition.
A measurement of these shifts at the LHC (and other colliders)
would therefore provide direct information about the nature 
of the electroweak phase transition.

  The outline of this papers is as follows.  
In Section~\ref{rates} we compute the effects of a light
stop on Higgs boson production and decay in the context 
of MSSM EWBG.  We discuss the implications of these results 
for Higgs boson searches at the Tevatron and LHC 
in Section~\ref{coll}.  Section~\ref{concl} is reserved for 
our conclusions.  Some additional discussion of parameter 
dependences is collected in Appendix~\ref{appa}.

\section{Gluon Fusion and Di-Photon Decays\label{rates}}

  The dominant Higgs boson production mode at the Tevatron
and the LHC is gluon fusion, 
$gg\to h^0$~\cite{Carena:2002es,Djouadi:2005gi}.  
In the SM, this process is dominated by a top quark loop.  
If the SM is extended to include new coloured particles coupling 
to the Higgs boson, such as stops in supersymmetry, they too will 
run in loops and contribute to the amplitude for this process.  
The interference with the top can be constructive 
or destructive, depending on the spin of the new particle 
and its couplings to the Higgs boson.

  For a lighter SM-like Higgs boson, with mass below about $130\,\gev$, 
the most effective discovery mode at the LHC is through its 
decays to photon pairs~\cite{Carena:2002es,Djouadi:2005gi}.  
This process arises 
from loops containing charged particles.  In the SM, the leading 
contribution to the amplitude comes from the $W^{\pm}$ gauge bosons, 
while the top quark provides a smaller contribution that interferes 
destructively.  When the SM is extended to include new charged states, 
these exotics will also contribute to the diphoton Higgs width.

  The Higgs sector of the MSSM is extended beyond the SM to include a pair of
$SU(2)_L$ doublets, and many new charged and colored states
couple to these doublets.  Even so, in much of the MSSM parameter space
consistent with current collider bounds, the phenomenology of the lightest
Higgs boson is very similar to that of the SM Higgs.  The direct Higgs
search bounds from LEP-II generally prefer larger values of the Higgs 
pseudoscalar mass parameter, $M_{A^0}\gg M_Z$, to push up the mass of 
the lightest $CP$-even Higgs boson $h^0$~\cite{Carena:2002es,Djouadi:2005gi}. 
In this limit, the couplings of the $h^0$ state to SM particles 
are nearly identical to those of the SM Higgs, up to corrections on 
the order of $M_Z^2/M_{A^0}^2$ from mixing with the other Higgs states, 
and loop effects due to the superpartners.  For superpartner
masses above a few hundred $\gev$, these loop corrections 
tend to be fairly mild, and the properties of the $h^0$ state
are very similar to those of the SM Higgs boson~\cite{Dawson:1996xz}.  

  In the small corner of the MSSM parameter space that 
is consistent with generating the baryon asymmetry through EWBG, 
there exists a very-light, mostly right-handed stop~\cite{Carena:2008vj}.
This state is unusual, compared to the other superpartners 
in this scenario, as well as the stops that are usually considered 
in other supersymmetric scenarios, in that it derives most of its mass 
from the vacuum expectation values~(VEVs) of the Higgs fields.  
As such, its effect on Higgs production and decay is potentially
comparable to that of the top quark.  It is precisely this effect that
we investigate in the present work.

  At tree-level in the MSSM, when the right-handed stop soft mass
$m_{U_3}^2$ is much smaller than the left-handed soft mass $m_{Q_3}^2$
and there is not much left-right stop mixing, the coupling of the lighter
mostly right-handed stop to the SM-like Higgs boson 
is given by~\cite{Carena:2008rt}
\beq
g_{h^0\tilde{t}_1\tilde{t}_1^*} \simeq \sqrt{2}\,v\,
\left[
|y_t|^2\,\sin^2\beta\,\left(1-\frac{|X_t|^2}{m_{Q_3}^2}\right) 
+ \frac{1}{3}\,g'^2\,\cos2\beta
\right],
\label{gthtree}
\eeq
where $X_t = (A_t - \mu\,\cot\beta)$, and we normalize 
$v=174\,\gev$.  When this coupling is positive, the light stop loops 
contributing to gluon fusion and di-photon decay interfere 
constructively with the loops of the top quark.
This coupling, when it is positive, is also proportional 
to the Higgs-stop interaction responsible for generating 
a strongly first-order electroweak phase transition in MSSM EWBG~\cite{
Cohen:1993nk,Carena:1996wj,Delepine:1996vn}.  
Thus, this scenario prefers weaker left-right stop mixing, 
and the increase in the strength of the phase transition is directly related 
to the effect of the light stop on the properties of the $h^0$ Higgs boson.
In contrast, when this coupling is negative or when there is 
strong left-right stop mixing (for which the expression of Eq.~\eqref{gthtree} 
is no longer valid) the interference of the stop loops with the top loops is 
destructive~\cite{Dermisek:2007fi}.~\footnote{The relevant expansion 
parameter for small left-right stop mixing is $m_tX_t/m_{Q_3}^2$.  
This quantity is always small in the light stop MSSM EWBG scenario, 
even for $|{X}_t|/m_{Q_3} \sim 1$, on account of the hierarchy 
$m_{t}/m_{Q_3} \lll 1$.}

  The $h^0\tilde{t}_1\tilde{t}_1^*$ coupling of Eq.~\eqref{gthtree} 
receives corrections enhanced by large logarithms of the ratio
$m_{Q_3}^2 /|m_{U_3}^2| \gg 1$  in the light stop scenario motivated 
by MSSM EWBG.  To re-sum these logs, the heavy SUSY and Higgs states 
should be integrated out at their mass thresholds~\cite{Carena:2008rt}.  
The resulting effective theory consists of the SM with a single Higgs 
boson $h^0$ augmented by a light stop $\tilde{t}_1$ and lighter 
gauginos and higgsinos.  
Equivalently, the effective theory coincides with the spectrum 
of focus--point~\cite{Feng:1999mn}, PeV-scale~\cite{Wells:2003tf},
or split supersymmetry~\cite{ArkaniHamed:2004fb} 
with an additional light stop.  
The matching and running conditions for this effective theory are 
computed in Ref.~\cite{Carena:2008rt} where all the heavy sfermions 
and heavy Higgs states are integrated out at a universal large 
mass scale $M$; $m_{Q_3}^2 = M^2 = M_A^2$, \emph{etc}.

\begin{figure}[ttt]
\vspace{1cm}
\begin{center}
        \includegraphics[width = 0.5\textwidth]{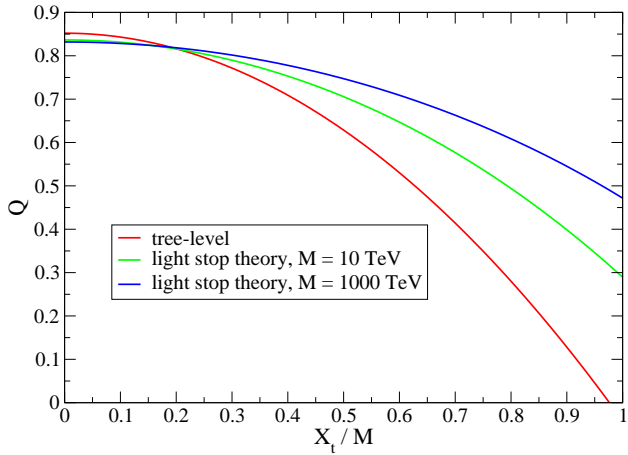}
\end{center}
\caption{Value of the low-energy Higgs-stop coupling $Q$ computed 
at tree-level, and using the light stop effective
theory.  The relevant model parameter values are taken to be
$m_{U_3}^2 = -(80 \,\gev)^2$, $\tan\beta = 10$,
and $M = m_{Q_3} = 10,\,1000\,\tev$.
}
\label{qcoupfig}
\end{figure}

  In this \emph{light-stop effective theory}~(LST), 
the coupling of the stop to the Higgs boson comes from the operator
\beq
\mathscr{L} \supset - Q\,|\tilde{t}_1^2||H|^2,
\label{qcoup}
\eeq
where $\tilde{t}_1$ is the light stop and $H = v+h^0/\sqrt{2}$ 
is the Higgs field.
The coupling $Q$ is obtained by matching at the scale $M$
where all the heavy Higgs bosons and sfermions are integrated out.
The matching condition is
\beq
Q(M) = g_{h^0\tilde{t}_1\tilde{t}_1^*}(M)/\sqrt{2}v
+(\mbox{threshold corrections}),
\label{gtheff}
\eeq
where $g_{h^0\tilde{t}_1\tilde{t}_1^*}$ is as in 
Eq.~\eqref{gthtree} with running parameters evaluated at 
scale $M$, and the threshold corrections are given 
in Ref.~\cite{Carena:2008rt}.  

The coupling relevant at lower scales, on the order of the Higgs mass,
is obtained by renormalization group~(RG) evolution in the effective
theory down from the scale $M$.
Expanding the Higgs field about its VEV, 
the interaction of Eq.~\eqref{qcoup}, evaluated near the Higgs boson
mass scale, then generates the effective $h^0\tilde{t}_1\tilde{t}_1^*$ 
coupling and contributes to the light stop mass according to
\bea
g_{h^0\tilde{t}_1\tilde{t}_1^*} &=& \sqrt{2}\,v\,Q,
\label{lststuff}\\
m_{\tilde{t}_1}^2 &=& m_{U_3}^2 + Q\,v^2.
\nnmb
\eeq
The coupling $Q$ is also responsible for driving the electroweak
phase transition to be first order when it is positive.
Therefore the effects of the light stop on the properties of
the Higgs boson are in direct proportion to the role of the
light stop in the electroweak phase transition.

  In Fig.~\ref{qcoupfig} we show the value of the coupling 
$Q$ at the light Higgs mass scale in the LST 
as well as the uncorrected tree-level value, 
$g_{h^0\tilde{t}_1\tilde{t_1}^*}/\sqrt{2}\,v$ given 
in Eq.~\eqref{gthtree}, evaluated using running parameters.  
We also set $m_{U_3}^2 = -(80\,\gev)^2$,
$\tan\beta = 10$, ${M} = 10,\,1000\,\tev$, and we assume 
the gluino is light and do not integrate it out explicitly.  
The tree-level and LST couplings are very similar for 
$|X_t|/M \sim 0$ but deviate significantly as this ratio grows, 
illustrating the improvement from the effective theory treatment.

  To compute the effect of the light stop on Higgs boson 
production and decay, we input the Higgs-stop coupling 
as well as the Higgs boson and light stop masses computed in 
the LST into CPSuperH~\cite{Lee:2003nta}.
We use a top quark mass of $m_t = 172.4\,\gev$, and the gaugino/higgsino
parameters $\mu = 190\,\gev$, $M_2 = 200\,\gev$, $M_1 = 100\,\gev$,
$M_3 = 700\,\gev$.
As in Ref.~\cite{Carena:2008rt}, we assume all sfermions
other than the light stop and all Higgs bosons other than
the $h^0$ are very heavy with a common mass $M$.
Unlike the sfermions, MSSM EWBG does not require that 
the heavy Higgs states be much heavier than the electroweak scale
with $M_A \simeq M$.
However, $M_A \gg M_Z$ helps to induce a strongly first-order
phase transition~\cite{Espinosa:1993yi}, and relaxing the assumption of 
$M_A \sim {M}$ will only modify our results by corrections 
on the order of $M_Z^2/M_A^2$~\cite{Carena:2002es,Djouadi:2005gi}.
In our treatment we also remove by hand the finite corrections
to the Yukawa couplings from heavy superpartners
since these decouple as $M\to \infty$
provided the gluino and higgsinos remain 
relatively light~\cite{Carena:2002bb}.

\begin{figure}[ttt]
\begin{minipage}[t]{0.47\textwidth}
        \includegraphics[width = \textwidth]{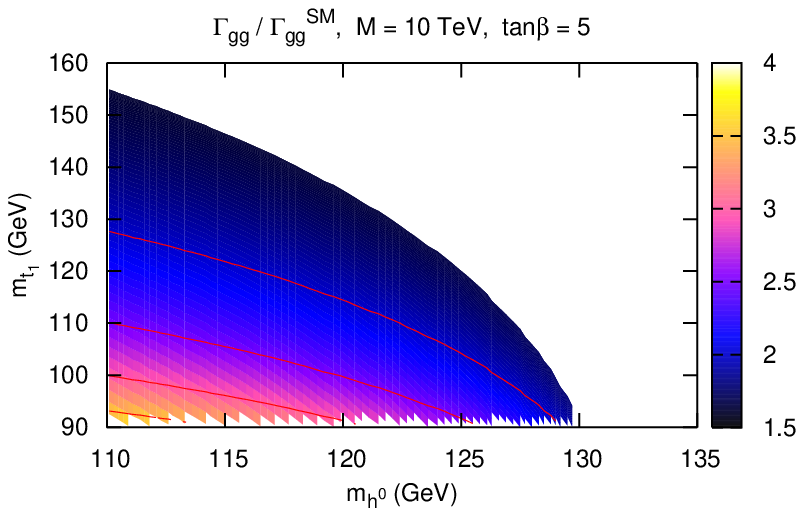}
\end{minipage}
\phantom{aa}
\begin{minipage}[t]{0.47\textwidth}
        \includegraphics[width = \textwidth]{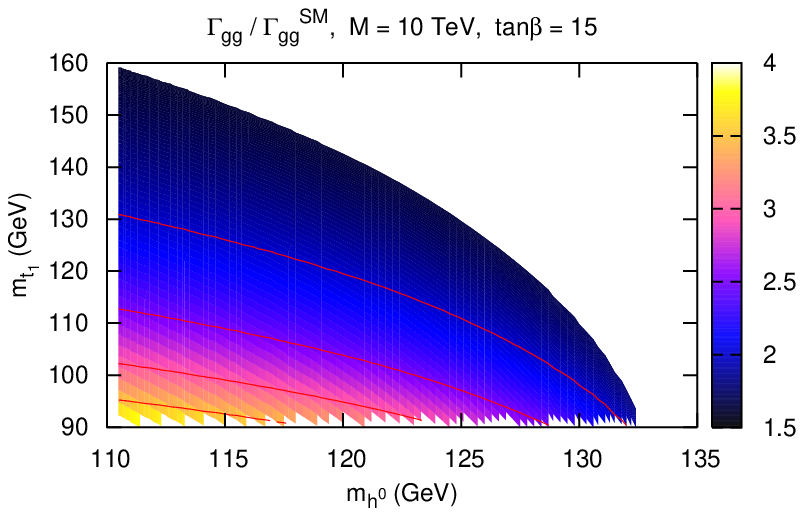}
\end{minipage}
\vspace{0.0cm}\\
\begin{minipage}[t]{0.47\textwidth}
        \includegraphics[width = \textwidth]{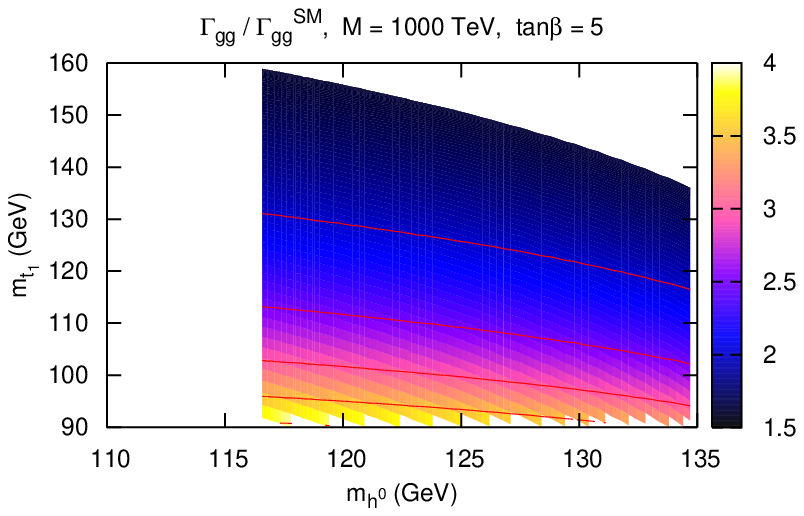}
\end{minipage}
\phantom{aa}
\begin{minipage}[t]{0.47\textwidth}
        \includegraphics[width = \textwidth]{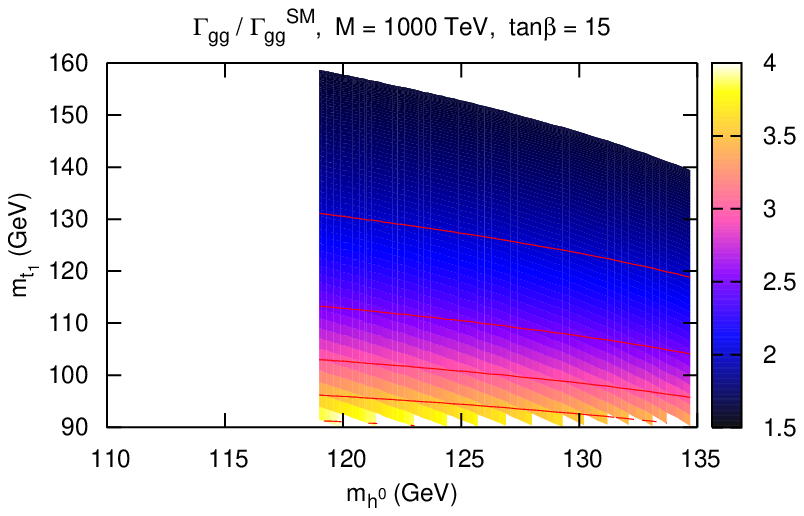}
\end{minipage}
\caption{Higgs boson decay width to gluons $\Gamma(h^0\to gg)$
relative to the SM as a function of $m_{h^0}$ and $m_{\tilde{t}_1}$
for ${M} = 10,\,1000\,\tev$ and $\tan\beta = 5,\,15$.
}
\label{hgg1}
\end{figure}

  In Fig.~\ref{hgg1} we show the ratio of the decay width 
of the Higgs boson to a pair of gluons
$\Gamma(h^0\to gg)$ computed in the LST relative to the 
value in the SM with the same value of the Higgs boson mass
as a function of $m_{h^0}$ and $m_{\tilde{t}_1}$
for $\tan\beta = 5,\,15$ and ${M} = 10,\,1000\,\tev$. 
In generating these figures we scan over the ranges
$-(150\gev)^2 \leq m_{U_3}^2 \leq (0\,\gev)^2$
and $0 \leq |{X}_t/{M}| \leq 0.9$.
These parameter ranges are a superset of the values 
that are consistent with a strongly first-order phase transition 
required for EWBG. 
(We also exhibit the dependence of the physical
masses $m_{\tilde{t}_1}$ and $m_{h^0}$ on the underlying Lagrangian
parameters in Appendix~\ref{appa}.)

  From Fig.~\ref{hgg1} we see a significant enhancement in the decay
width to gluons relative to the SM by as much as a factor of four.
Since this decay width is nearly proportional
to the Higgs production rate through gluon fusion at hadron colliders
at leading order~(LO), including the Tevatron at $\sqrt{s} = 1.96\,\tev$
and the LHC at $\sqrt{s} = 10\!-\!14\tev$,
our results imply a strong enhancement in this production mode.\footnote{
Our computation of the $\Gamma(h^0\to gg)$ width is at LO.  
While NLO corrections are significant, their effect is to rescale both 
the SM and squark LO contributions to the production rate in nearly the 
same way~\cite{Dawson:1996xz,Harlander:2004tp,Degrassi:2008zj}.  
Thus, we expect 
the bulk of these higher-order corrections to cancel in the ratios 
of decay widths, interpreted as ratios of gluon fusion production rates,
that we display.}
The enhancement is greatest for smaller values of 
the Higgs boson $m_{h^0}$ and stop $m_{\tilde{t}_1}$ masses,
corresponding to smaller values of $m_{U_3}^2$ and $|{X}_t|/m_{Q_3}$.
Indeed, it is for these smaller mass values that the electroweak
phase transition can be strong enough to allow viable EWBG.
The recent analysis of Ref.~\cite{Carena:2008vj} finds that this region
falls within the lower-left corner of the $m_h^0$--$m_{\tilde{t}_1}$
plane with  $m_{\tilde{t}_1},\,m_{h^0} < 125\,\gev$,
where the enhancement is greater than a factor of two.

\begin{figure}[ttt]
\begin{minipage}[t]{0.47\textwidth}
        \includegraphics[width = \textwidth]{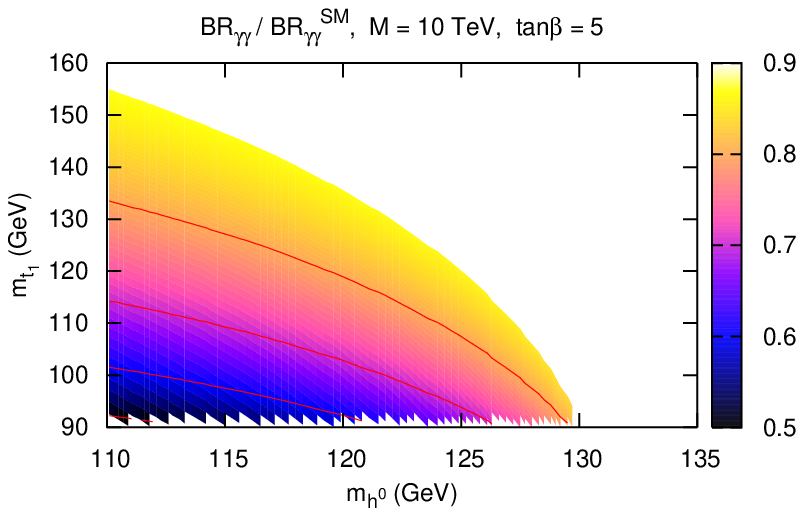}
\end{minipage}
\phantom{aa}
\begin{minipage}[t]{0.47\textwidth}
        \includegraphics[width = \textwidth]{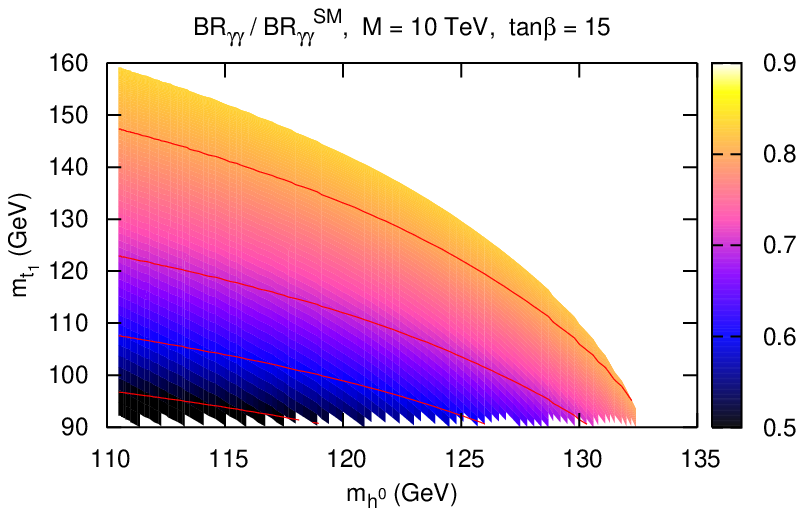}
\end{minipage}
\vspace{0.0cm}\\
\begin{minipage}[t]{0.47\textwidth}
        \includegraphics[width = \textwidth]{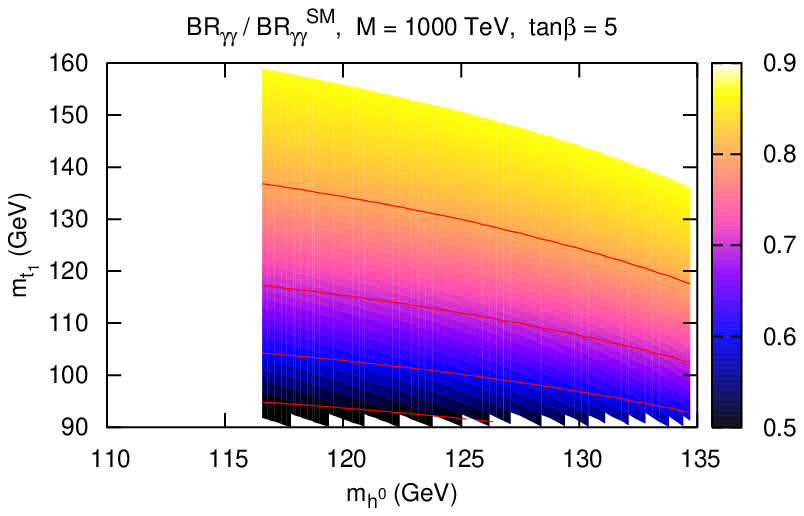}
\end{minipage}
\phantom{aa}
\begin{minipage}[t]{0.47\textwidth}
        \includegraphics[width = \textwidth]{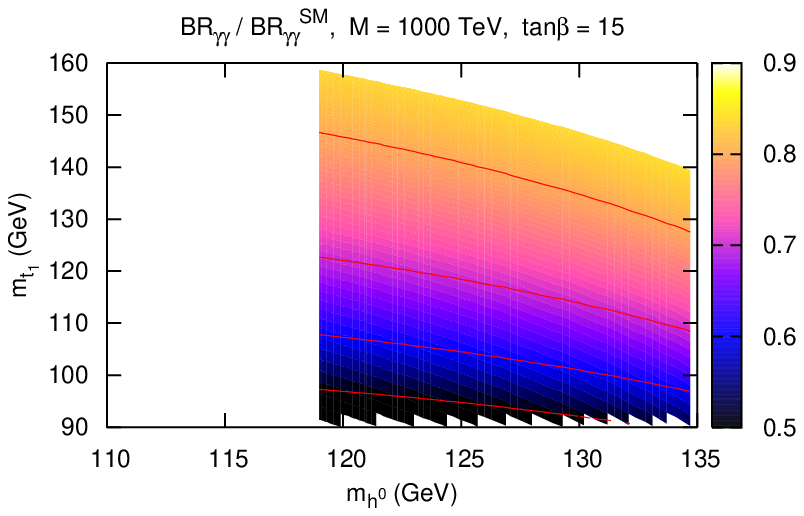}
\end{minipage}
\caption{Higgs boson branching fraction to diphotons 
$BR(h^0\to \gamma\gamma)$ relative to the SM as a function 
of $m_{h^0}$ and $m_{\tilde{t}_1}$ for ${M} = 10,\,1000\,\tev$ 
and $\tan\beta = 5,\,15$.
}
\label{hgg2}
\end{figure}

  We show in Fig.~\ref{hgg2} the value of the branching fraction
$BR(h^0\to\gamma\gamma)$ in the light stop scenario relative to
the SM for the same value of the Higgs boson mass as a function of 
$m_{h^0}$ and $m_{\tilde{t}_1}$.
As above, we consider $\tan\beta = 5,\,15$ and ${M} = 10,\,1000\,\tev$,
and scan over the ranges $-(150\gev)^2 \leq m_{U_3}^2 \leq (0\,\gev)^2$
and $0 \leq |{X}_t/{M}| \leq 0.9$.
The diphoton branching fraction is significantly suppressed,
particularly for the lower values of $m_{\tilde t}$ and $m_{h^0}$
favoured by EWBG.  The suppression originates from two sources.
First, there is destructive interference between the stop and
the $W^{\pm}$ gauge bosons in the loop-level amplitude 
for $h^0\to \gamma\gamma$.  Second, the enhancement of the Higgs boson
decay width into gluons increases the total decay width, thereby
diluting the fraction of decays to photon pairs.
Between these two effects, the suppression factor for  
$BR(h^0\to \gamma\gamma)$ relative to the SM is always less about 0.7
and as small as 0.5 within the region of parameters consistent
with EWBG.  The light charginos that are also
required for successful EWBG can further modify the Higgs boson 
decay width to diphotons.  We find that this effect is less than
about $5\%$ once the LEP-II bound of $104\,\gev$ is imposed on
the lightest chargino~\cite{Abbiendi:2003sc}.

\begin{figure}[ttt]
\begin{minipage}[t]{0.47\textwidth}
        \includegraphics[width = \textwidth]{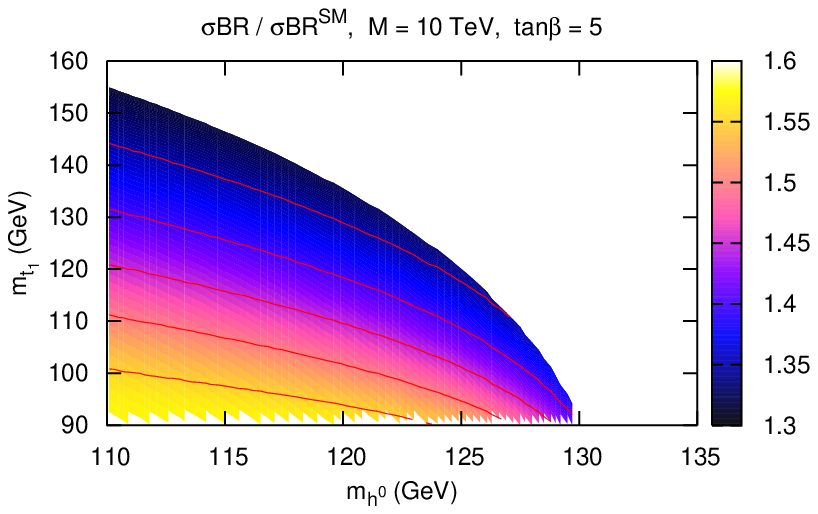}
\end{minipage}
\phantom{aa}
\begin{minipage}[t]{0.47\textwidth}
        \includegraphics[width = \textwidth]{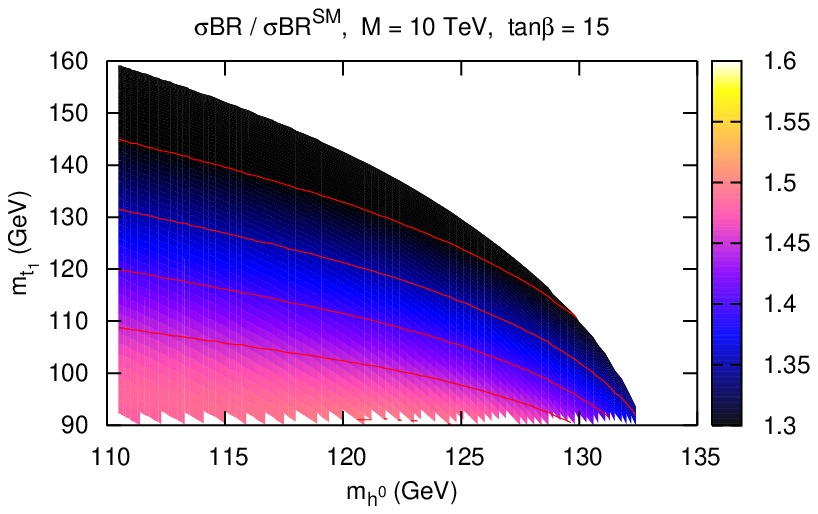}
\end{minipage}
\vspace{0.0cm}\\
\begin{minipage}[t]{0.47\textwidth}
        \includegraphics[width = \textwidth]{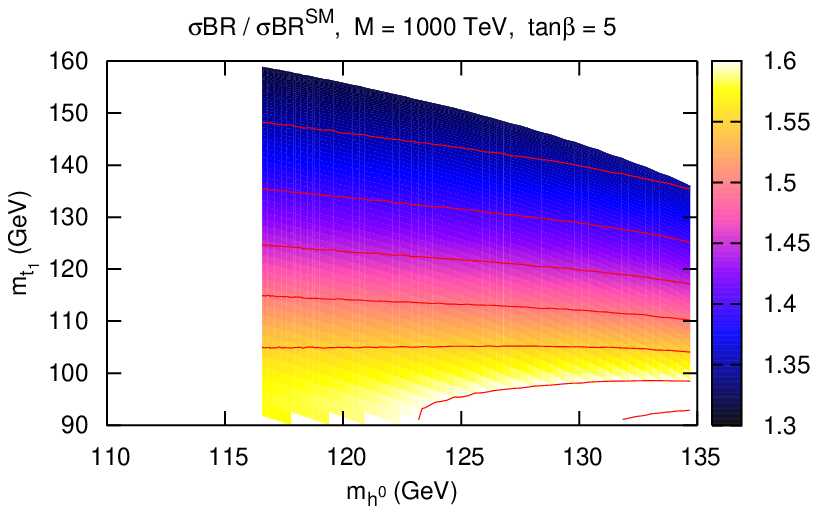}
\end{minipage}
\phantom{aa}
\begin{minipage}[t]{0.47\textwidth}
        \includegraphics[width = \textwidth]{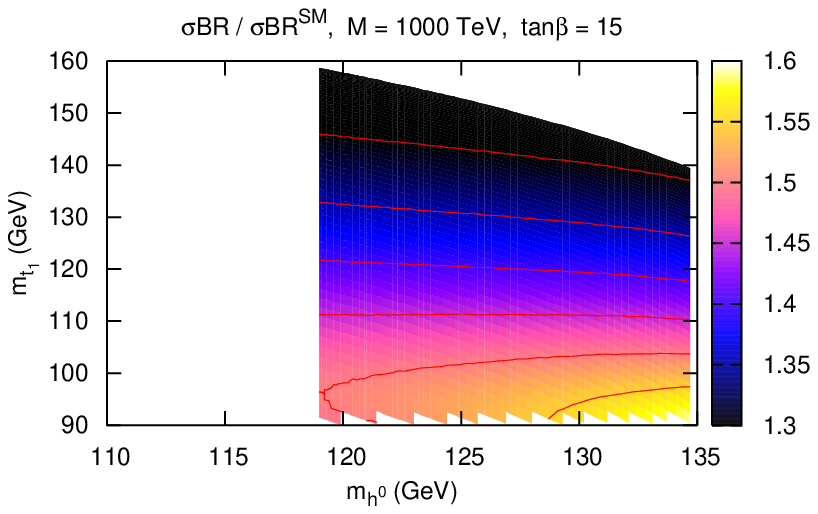}
\end{minipage}
\caption{Inclusive $pp\to h^0\to \gamma\gamma$ production rate 
at the LHC relative to the SM as a function of $m_{h^0}$ and 
$m_{\tilde{t}_1}$ for ${M} = 10,\,1000\,\tev$ and $\tan\beta = 5,\,15$.  
}
\label{hgg3}
\end{figure}

  In Fig.~\ref{hgg3} we show contours of the total inclusive
$pp\to h^0 \to \gamma\gamma$ production rate in the light stop scenario
relative to the SM with the same value of the Higgs boson mass 
as a function of $m_{h^0}$ and $m_{\tilde{t}_1}$.
As above, we consider $\tan\beta = 5,\,15$ and ${M} = 10,\,1000\,\tev$,
and scan over the ranges $-(150\gev)^2 \leq m_{U_3}^2 \leq (0\,\gev)^2$
and $0 \leq |{X}_t/{M}| \leq 0.9$.
We also assume that gluon fusion makes up $83\%$ of the inclusive 
production rate before including the enhancement from a light stop, 
which is approximately the expected fraction contributing to the 
inclusive signal for a light SM Higgs boson at the LHC with 
$\sqrt{s} = 14\,\tev$~\cite{Ball:2007zza,Aad:2009wy}.
From Figs.~\ref{hgg1} and \ref{hgg2} we know that the Higgs boson
production rate through gluon fusion is enhanced while the branching 
ratio into diphotons is suppressed.  The total rate is approximately 
proportional to the product of these quantities.  This product
is ultimately enhanced in the light stop scenario because the
stop loop interferes constructively with the top quark loop in the production
rate and destructively with a more dominant $W^{\pm}$ loop 
in the decay width to diphotons.  In the region of parameter
space consistent with a strongly first-order phase transition,
the inclusive production rate is enhanced by a factor between 
$1.4$ and $1.6$.

  The light stop in the MSSM EWBG scenario will also lead to
modifications of other Higgs boson search channels.
There will be an enhancement in all channels for which gluon 
fusion is the dominant production mechanism.  For example,
the rate for inclusive $pp \to h^0 \to ZZ^*$ at the LHC
(assuming the gluon fusion makes up $83\%$ of the total rate)
is increased by a factor of 1.75--3 within the parameter region 
consistent with viable EWBG.  On the other hand, the suppression 
in the $h^0\to \gamma\gamma$ branching fraction reduces proportionally
the signal rates from Higgs production through vector-boson-fusion~(VBF)
or in association with $t\bar{t}$ or $W/Z$.  Interesting further
possibilities that we do not explore are Higgs boson production in
association with light stops~\cite{Djouadi:1997xx,Dedes:1999ku},
or a modification of the Higgs self-couplings~\cite{Noble:2007kk}.

\section{Prospects at the Tevatron and LHC \label{coll}}

  The results of the previous section indicate that within the parameter
region consistent with MSSM EWBG the production and decay of modes 
of the $h^0$ Higgs boson are significantly modified relative
to the SM by the presence of a light stop.  Measuring Higgs boson
couplings  at the Tevatron and LHC will be extremely 
challenging, and will require both a large amount of integrated luminosity 
and a detailed understanding of systematics. Even so, the effects of a 
light stop consistent with MSSM EWBG on $h^0$ Higgs boson observables 
are large enough that they can likely be distinguished from a SM Higgs boson 
by LHC data, and may also have a discernible impact on Tevatron Higgs searches.

  Before discussing the future prospects of detecting this scenario, 
let us also point out that the modifications of the Higgs properties 
induced by a light stop do not run afoul of existing collider bounds 
or modify the limits imposed by previous Higgs boson searches.  
The most stringent bound comes from LEP-II, which places a lower
limit on the SM Higgs boson mass of $114.7\,\gev$~\cite{Amsler:2008zzb}.
This limit also applies to the $h^0$ state in the light stop scenario.

  Present limits from the Tevatron strongly constrain 
a SM Higgs boson with mass near $160\,\gev$ through searches 
for inclusive Higgs production with decays to $W^+W^-$~\cite{cdfreach}.  
For lighter SM-like Higgs bosons with $m_{H^0}\lesssim 125\,\gev$,
the Tevatron bounds are much weaker and are dominated by
the $W/Z$ associated production channels with decays to $b\bar{b}$.
These will not be significantly modified by the presence of 
a light stop.  Among the inclusive search channels (dominated by gluon fusion) 
that will be enhanced by a light stop, the most promising is $h^0\to WW^*$.  
Current Tevatron data with $3.0\,fb^{-1}$ of integrated luminosity
constrain the total rate for this mode, for $m_{h^0}\lesssim 125\,\gev$,
to be less than about $8$ times that predicted by the SM~\cite{
Aaltonen:2008ec,Abazov:2009kq}.
We find the enhancement in the light stop scenario consistent with
EWBG to be in the range of 2--4 times the SM.  However, with the
projected integrated luminosity on the order of $10\,fb^{-1}$
and expected improvements in Higgs search analyses~\cite{cdfreach},
the Tevatron may potentially be able to probe the enhancement in
inclusive Higgs production due to a light stop through this 
channel, particularly for $h^0$ masses towards
the upper range ($\sim 125\,\gev$) of what is consistent with EWBG.

  At the LHC, the most effective search mode for a light SM-like 
Higgs boson with $m_{h^0} \lsim 135\,\gev$ is inclusive
$h^0 \to \gamma\gamma$~\cite{Ball:2007zza,Aad:2009wy}.  
This channel also allows for a precise measurement of the Higgs boson 
mass with an uncertainty below $0.2\,\gev$ with $30\,fb^{-1}$ of 
data~\cite{Ball:2007zza}.
Inclusive $h^0\to ZZ^*\to 4\ell$, as well as $h^0 \to \gamma\gamma$ and
$h^0 \to \tau\tau$ through vector boson fusion~(VBF), 
and $h^0\to \gamma\gamma$ via production in association with $W/Z$
or $t\bar{t}$ are also relevant with a large integrated 
luminosity~\cite{Ball:2007zza,Aad:2009wy}.
Including the effects of a light stop, the inclusive $h^0 \to \gamma\gamma$
and $h^0\to ZZ^*$ channels are significantly enhanced, while the net rates
for $h^0 \to \gamma\gamma$ through VBF or associated production 
are somewhat suppressed.  By comparing the rates for these various
channels, it may be possible to detect the enhancement in the gluon
fusion rate from a light stop.

  A program to extract Higgs boson couplings from LHC data was
outlined in Refs.~\cite{Zeppenfeld:2000td,Belyaev:2002ua,Duhrssen:2004cv}.
The estimated rates and systematic uncertainties 
(mostly from higher order corrections, PDFs, and luminosity) 
used in Refs.~\cite{Zeppenfeld:2000td,Belyaev:2002ua,Duhrssen:2004cv}
stand up quite well to the more recent analyses of 
Refs.~\cite{Ball:2007zza,Aad:2009wy} with the exception of the 
$t\bar{t}h \to t\bar{t}b\bar{b}$ channel which does not play an
important role in extracting the Higgs decay width to gluons or photons.
Refs.~\cite{Zeppenfeld:2000td,Belyaev:2002ua,Duhrssen:2004cv} find 
that the partial width  $\Gamma(h^0\to gg)$ can be determined for a SM
Higgs boson with a 1$\sigma$ error of 30\% (90\%) with $200\,fb^{-1}$ 
($30\,fb^{-1}$) of integrated luminosity.  This estimate is obtained primarily 
from a comparison of the inclusive $h^0\to \gamma\gamma$ rate to that from VBF,
and assumes SM-strength Higgs boson couplings to the electroweak gauge bosons.

  The analysis of Refs.~\cite{Zeppenfeld:2000td,Belyaev:2002ua,Duhrssen:2004cv}
can be applied directly to the light stop scenario,
for which the assumption of SM-strength electroweak gauge boson couplings
holds to an excellent approximation.  The reduction in the VBF di-photon 
signal will increase the statistical error on the determination of 
$\Gamma(h^0\to gg)$, but the effect will be small (for $200\,fb^{-1}$
of data) relative to the assumed $20\%$ systematic uncertainty on
the (SM) gluon fusion rate.  Thus, we expect an uncertainty on the
gluon width of about $30\%$ with $200\,fb^{-1}$ of data.  
Given the light stop enhancement of this width by a factor
of at least two in the region consistent with EWBG, the enhancement in the 
gluon decay width relative to the SM can be detected at the LHC with a 
significance greater than $3\sigma$.

  A similar estimate can be obtained using the updated ATLAS detector-level 
analysis of Ref.~\cite{Aad:2009wy}.  Here, the comparison would be
between the rates for the inclusive $h^0\to \gamma\gamma$ channel 
and the di-photon channels in association with one or two hard jets.  
The inclusive channel is dominated by gluon fusion, while the channels
involving additional jets receive a larger contribution from VBF.  
For a light stop in the MSSM EWBG region, the enhancements in these
channels relative to the SM are by factors of about 
$1.6,\,1.2,\,0.7$ respectively. 
Based on statistics alone, we find that it will be possible to easily 
distinguish this pattern from the SM with $200\,fb^{-1}$ of data.
To be effective when systematics are included, however, a better
 understanding of the gluon fusion contribution to the two-jet 
channel is needed~\cite{Aad:2009wy}.

  It will be more challenging to directly probe the 
effect of the light stop on $\Gamma(h^0\to \gamma\gamma)$.  
Refs.~\cite{Zeppenfeld:2000td,Belyaev:2002ua,Duhrssen:2004cv} 
find that this width can be determined for a SM Higgs boson with 
an error of 20\% (40\%) with $200$~fb$^{-1}$ 
($30\,fb^{-1}$) of data.
With a light stop, the reduction in $BR(h^0\to \gamma\gamma)$ will further 
degrade the statistics in the non-gluon fusion production modes.  
Thus, it does not appear to be possible to see the stop effects 
on $\Gamma(h^0\to \gamma\gamma)$ above the $2\sigma$ level.
However, relative to Refs.~\cite{Zeppenfeld:2000td,Belyaev:2002ua,
Duhrssen:2004cv}, a very light stop enhances the prospects 
for $h^0\to ZZ^*\to 4\ell$. The ratio of this rate to that for inclusive 
$h^0\to \gamma\gamma$ may allow for an improved test of the
effect of the light stop on $\Gamma(h^0\to \gamma\gamma)$
for Higgs boson masses towards the upper end of the range consistent
with MSSM EWBG, although it will be limited by statistics.  

  Let us emphasize that the estimates made above based on 
Refs.~\cite{Zeppenfeld:2000td,Belyaev:2002ua,Duhrssen:2004cv} are 
conservative, in that improvements in the systematic uncertainties 
associated with Higgs boson production and decay at the LHC are likely
to further improve the determination of Higgs couplings.  
For example, Refs.~\cite{Zeppenfeld:2000td,Belyaev:2002ua,Duhrssen:2004cv}
assumed a $20\%$ error in the prediction for the SM gluon fusion rate.
Significant progress has been made recently in computing this
rate at higher orders with the inclusion of electroweak 
corrections~\cite{Actis:2008ts,Anastasiou:2008tj,deFlorian:2009hc}, 
along with a re-summation of the apparent leading 
loop-level enhancements~\cite{Ahrens:2008qu}.  
Together, these indicate perturbative uncertainty less 
than $3\%$, down from about $10\%$~\cite{Ahrens:2008qu}.
There is an additional estimated $10\%$ uncertainty from the 
parton distribution functions~\cite{Djouadi:2003jg}, 
which could potentially 
be reduced with LHC data~\cite{Stirling:2008sj,Dittmar:2009ii}.  
Along with a large amount of luminosity,
these and future advances may make it possible to observe 
the effect of a light stop on $\Gamma(h^0\to \gamma\gamma)$, 
and to even estimate the Higgs-stop coupling $Q$ within the 
context of this scenario.  A reduction in the various systematic
uncertainties would also greatly improve the ability of 
the SLHC to probe Higgs boson couplings~\cite{Gianotti:2002xx}.

\section{Conclusions\label{concl}}

  MSSM EWBG can account for the baryon asymmetry of the universe
provided the lightest CP-even Higgs boson $h^0$ is SM-like and 
there exists a mostly right-handed stop that is significantly
lighter than the top quark.  In the present work we have investigated 
the effect of this light stop on $h^0$ Higgs boson production and decay.
We find a significant enhancement in the Higgs production rate through
gluon fusion and a less pronounced suppression of the Higgs boson branching
fraction to pairs of photons.  The enhancement in $\Gamma(h^0\to gg)$
is large enough that it can potentially be detected at the LHC after
several years of running.  

  Similar enhancements of the gluon fusion rate can arise in a variety
of contexts, such as with fourth-generation~\cite{Kribs:2007nz,
Arik:2007vi,Fok:2008yg} 
or exotic quarks~\cite{Morrissey:2003sc,Ham:2008xf}.  On the other hand,
the rate for gluon fusion is suppressed within the \emph{golden region}
of the MSSM where fine-tuning is minimized~\cite{Dermisek:2007fi,Low:2009nj},
as well as in many little Higgs models~\cite{Han:2003gf}, and in other
contexts~\cite{Belyaev:2005ct,Cacciapaglia:2009ky,
Barger:2009me,Bhattacharyya:2009nb}.  
Within the MSSM, the observation of an enhancement
in the gluon fusion Higgs production rate would provide evidence for
a light stop that is complementary to direct collider searches 
for this state.  While these direct search channels can provide a more
efficient stop discovery mode, they often do not yield much information
about the nature of the electroweak phase transition or the composition
of the light stop state.  The observation of direct stop signals
in combination with modified Higgs boson production and decay 
signatures would together provide evidence in favour of MSSM EWBG
and a direct test of the strength of the electroweak phase transition.

\section*{Acknowledgements}

We thank Marcela Carena, Sally Dawson, Sven Heinemeyer, Gudrun Hiller, 
Gordy Kane, Germano Nardini, Aaron Pierce, Marco Pieri, Matt Schwartz, 
Carlos Wagner, Lian-Tao Wang, and James Wells for helpful 
comments and discussions.  We also thank Carlos Wagner and James
Wells for reading an early version of the manuscript.
This work was supported by the Harvard Center for the Fundamental
Laws of Nature, Kavli Institute for Theoretical Physics~(KITP),
the Michigan Center for Theoretical Physics,
as well as the DOE under Grant DE-FG02-95ER40899 and the
National Science Foundation under Grant No.~PHY05-51164.
D.M. also acknowledges CERN-TH for their hospitality.


\appendix

\section{Appendix: Higgs and Stop Mass Dependence\label{appa}}

\begin{figure}[ttt]
\begin{minipage}[t]{0.47\textwidth}
        \includegraphics[width = \textwidth]{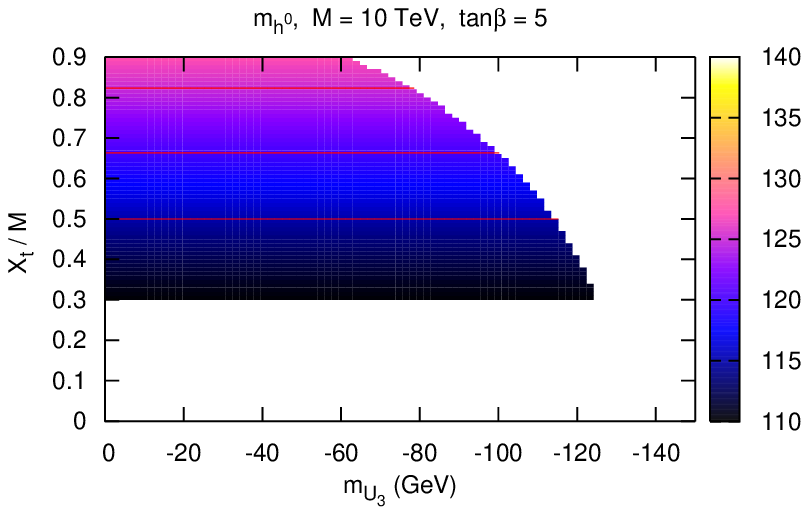}
\end{minipage}
\phantom{aa}
\begin{minipage}[t]{0.47\textwidth}
        \includegraphics[width = \textwidth]{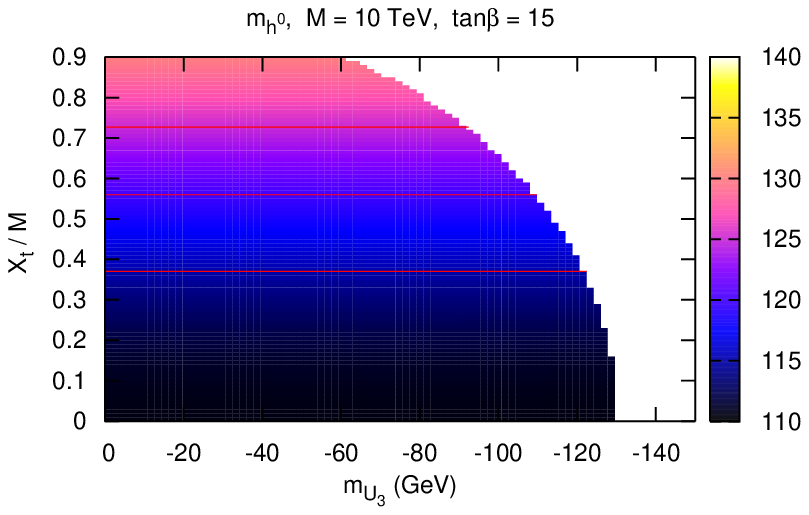}
\end{minipage}
\vspace{0.0cm}\\
\begin{minipage}[t]{0.47\textwidth}
        \includegraphics[width = \textwidth]{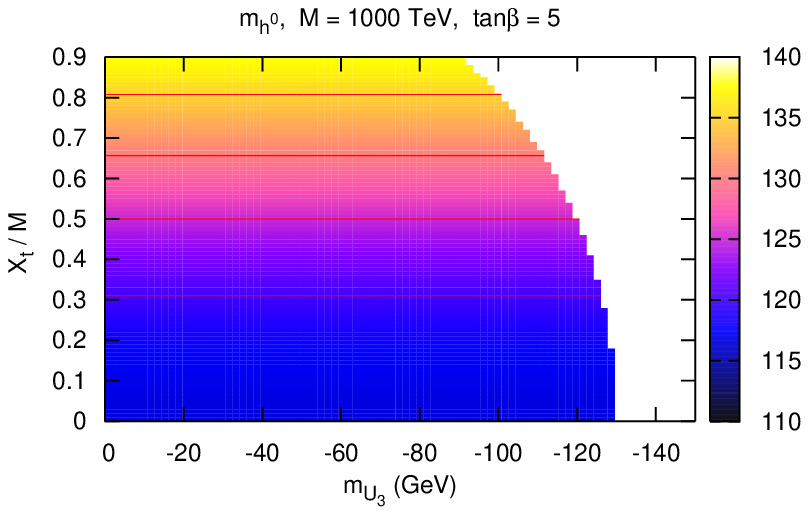}
\end{minipage}
\phantom{aa}
\begin{minipage}[t]{0.47\textwidth}
        \includegraphics[width = \textwidth]{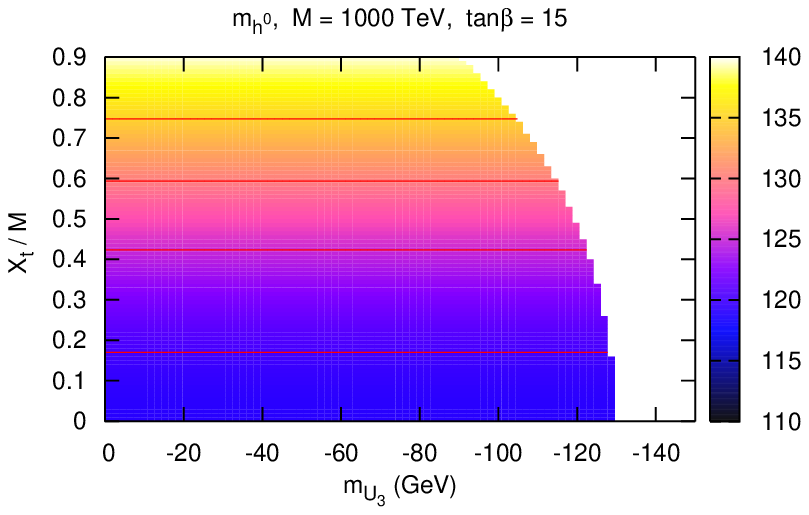}
\end{minipage}
\caption{Higgs boson masses as a function of $m_{U_3}$ and
$|X_t|/M$ for $\tan\beta = 5,\,15$ and $M = 10,\,1000\,\tev$.
}
\label{hm1}
\end{figure}

\begin{figure}[htt]
\begin{minipage}[t]{0.47\textwidth}
        \includegraphics[width = \textwidth]{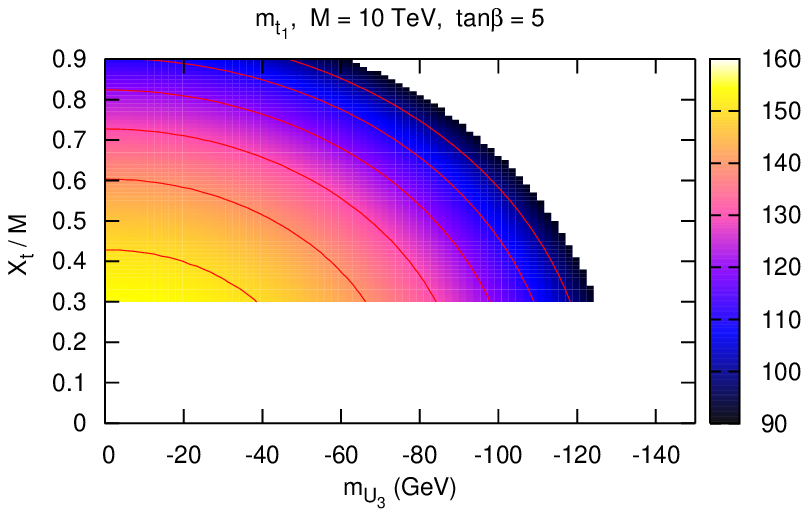}
\end{minipage}
\phantom{aa}
\begin{minipage}[t]{0.47\textwidth}
        \includegraphics[width = \textwidth]{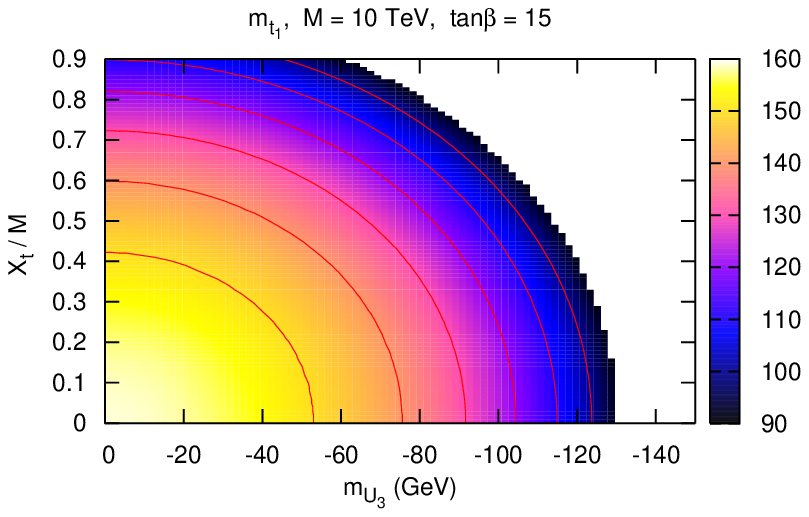}
\end{minipage}
\vspace{0.0cm}\\
\begin{minipage}[t]{0.47\textwidth}
        \includegraphics[width = \textwidth]{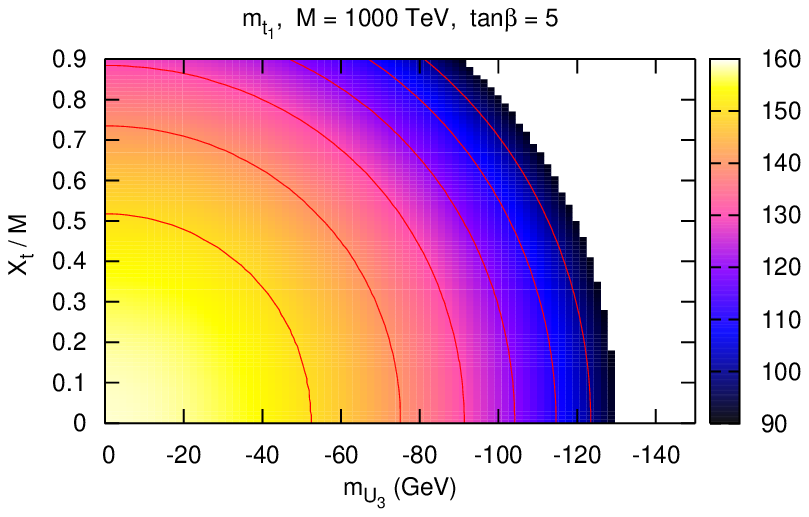}
\end{minipage}
\phantom{aa}
\begin{minipage}[t]{0.47\textwidth}
        \includegraphics[width = \textwidth]{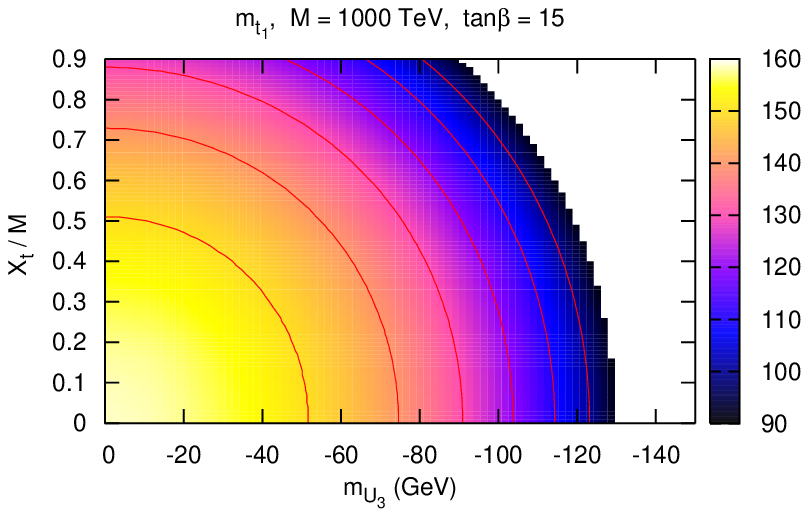}
\end{minipage}
\caption{Light stop masses as a function of $m_{U_3}$ and
$|X_t|/M$ for $\tan\beta = 5,\,15$ and $M = 10,\,1000\,\tev$.
}
\label{hm2}
\end{figure}

  We collect here several plots of $h^0$ Higgs boson and $\tilde{t}_1$
stop masses as functions of the underlying MSSM parameters.
In Fig.~\ref{hm1} we show contours of the Higgs boson mass
as a function of the underlying parameters $m_{U_3}$ and
$|X_t|/M$ for $\tan\beta = 5,\,15$ and $M = 10,\,1000\,\tev$
computed at one-loop order in the LST.
In generating these figures we scan over the ranges
$-(150\gev)^2 \leq m_{U_3}^2 \leq (0\,\gev)^2$
and $0 \leq |{X}_t/{M}| \leq 0.9$, and use a top quark mass
of $m_t =  172.4\,\gev$. 
The unfilled regions of these plots correspond to areas where 
$m_{h^0}< 110\,\gev$ or $m_{\tilde{t}_1} < 90\,\gev$.
These limits are slightly weaker than the current LEP-II 
mass bounds: $m_{h^0} > 114.7\,\gev$, 
and $m_{\tilde{t}_1} > 95.7 \,\gev$~\cite{Amsler:2008zzb}.
We apply weaker bounds here for illustrative purposes,
to account for theoretical uncertainties (particularly
in the $h^0$ Higgs mass), and for the weakening of the 
stop mass bound for stop-LSP mass differences less than $10\,\gev$.
>From Fig.~\ref{hm1} we observe that for $M =1000$~TeV the Higgs boson
mass is $\gsim 118$~GeV for these ranges of $X_t$ and $m_{U_3}^2$. 
We also see that the mass falls for smaller values of $|X_t|$,
but is relatively insensitive to value of $m_{U_3}^2$. 

  We show contours of the light $\tilde{t}_1$ stop mass
in Fig.~\ref{hm2} for the same parameter ranges of
$X_t$ and $m_{U_3}^2$ as were used in Fig.~\ref{hm1}.
Again, the unfilled regions of these plots correspond
to areas where $m_{h^0}< 110\,\gev$ or $m_{\tilde{t}_1} < 90\,\gev$.
At lower values of $X_t$ and $|m_{U_3}^2|$ the stop mass is close 
to that of the top quark, and its value decreases with increasing 
values of $X_t$ and $-m_{U_3}^2$.




\begin{thebibliography}{9}

\bibitem{Martin:1997ns}
For phenomenological reviews of supersymmetry see:\\
  S.~P.~Martin,
  [hep-ph/9709356];
  D.~J.~H.~Chung, L.~L.~Everett, G.~L.~Kane, S.~F.~King, J.~D.~Lykken and L.~T.~Wang,
  Phys.\ Rept.\  {\bf 407}, 1 (2005)
  [hep-ph/0312378].

\bibitem{Kuzmin:1985mm}
  V.~A.~Kuzmin, V.~A.~Rubakov and M.~E.~Shaposhnikov,
  Phys.\ Lett.\  B {\bf 155}, 36 (1985);
  M.~E.~Shaposhnikov,
  Nucl.\ Phys.\  B {\bf 299}, 797 (1988);
  M.~E.~Shaposhnikov,
  Nucl.\ Phys.\  B {\bf 287}, 757 (1987).


\bibitem{Cohen:1993nk}
For reviews of (electroweak) baryogenesis see:\\
  A.~G.~Cohen, D.~B.~Kaplan and A.~E.~Nelson,
  Ann.\ Rev.\ Nucl.\ Part.\ Sci.\  {\bf 43}, 27 (1993)
  [hep-ph/9302210];
  V.~A.~Rubakov and M.~E.~Shaposhnikov,
  Usp.\ Fiz.\ Nauk {\bf 166}, 493 (1996)
  [Phys.\ Usp.\  {\bf 39}, 461 (1996)]
  [hep-ph/9603208];
  M.~Trodden,
  Rev.\ Mod.\ Phys.\  {\bf 71}, 1463 (1999)
  [hep-ph/9803479].
%
  A.~Riotto,
  [hep-ph/9807454];
  M.~Quiros,
  [hep-ph/9901312];
  A.~Riotto and M.~Trodden,
  Ann.\ Rev.\ Nucl.\ Part.\ Sci.\  {\bf 49}, 35 (1999)
  [hep-ph/9901362];
  J.~M.~Cline,
  Pramana {\bf 55}, 33 (2000)
  [hep-ph/0003029].
  J.~M.~Cline,
  [hep-ph/0609145].


\bibitem{Gavela:1993ts}
  M.~B.~Gavela, P.~Hernandez, J.~Orloff and O.~Pene,
  Mod.\ Phys.\ Lett.\  A {\bf 9}, 795 (1994)
  [hep-ph/9312215];
  M.~B.~Gavela, M.~Lozano, J.~Orloff and O.~Pene,
  Nucl.\ Phys.\  B {\bf 430}, 345 (1994)
  [hep-ph/9406288];
  M.~B.~Gavela, P.~Hernandez, J.~Orloff, O.~Pene and C.~Quimbay,
  Nucl.\ Phys.\  B {\bf 430}, 382 (1994)
  [hep-ph/9406289].

\bibitem{Huet:1994jb}
  P.~Huet and E.~Sather,
  Phys.\ Rev.\  D {\bf 51}, 379 (1995)
  [hep-ph/9404302].

\bibitem{RamseyMusolf:2006vr}
  M.~J.~Ramsey-Musolf and S.~Su,
  Phys.\ Rept.\  {\bf 456}, 1 (2008)
  [hep-ph/0612057].

\bibitem{Carena:1997gx}
  M.~S.~Carena, M.~Quiros, A.~Riotto, I.~Vilja and C.~E.~M.~Wagner,
  Nucl.\ Phys.\  B {\bf 503}, 387 (1997)
  [hep-ph/9702409].
  M.~S.~Carena, J.~M.~Moreno, M.~Quiros, M.~Seco and C.~E.~M.~Wagner,
  Nucl.\ Phys.\  B {\bf 599}, 158 (2001)
  [hep-ph/0011055];
  M.~S.~Carena, M.~Quiros, M.~Seco and C.~E.~M.~Wagner,
  Nucl.\ Phys.\  B {\bf 650}, 24 (2003)
  [hep-ph/0208043].

\bibitem{Multamaki:1997ep}
  T.~Multamaki and I.~Vilja,
  Phys.\ Lett.\  B {\bf 411}, 301 (1997)
  [hep-ph/9705469].


\bibitem{Cline:1997vk}
  J.~M.~Cline, M.~Joyce and K.~Kainulainen,
  Phys.\ Lett.\  B {\bf 417}, 79 (1998)
  [Erratum-ibid.\  B {\bf 448}, 321 (1999)]
  [hep-ph/9708393].
  J.~M.~Cline and K.~Kainulainen,
  Phys.\ Rev.\ Lett.\  {\bf 85}, 5519 (2000)
  [hep-ph/0002272].
  J.~M.~Cline, M.~Joyce and K.~Kainulainen,
  JHEP {\bf 0007}, 018 (2000)
  [hep-ph/0006119].
  J.~M.~Cline, M.~Joyce and K.~Kainulainen,
  [hep-ph/0110031].

\bibitem{Riotto:1997gu}
  A.~Riotto,
  Int.\ J.\ Mod.\ Phys.\  D {\bf 7}, 815 (1998)
  [hep-ph/9709286].

\bibitem{Delepine:2002as}
  D.~Delepine, R.~Gonzalez Felipe, S.~Khalil and A.~M.~Teixeira,
  Phys.\ Rev.\  D {\bf 66}, 115011 (2002)
  [:hep-ph/0208236].



\bibitem{Abel:2001vy}
See for example:\\
  S.~Abel, S.~Khalil and O.~Lebedev,
  Nucl.\ Phys.\  B {\bf 606}, 151 (2001)
  [hep-ph/0103320].



\bibitem{Murayama:2002xk}
  H.~Murayama and A.~Pierce,
  Phys.\ Rev.\  D {\bf 67}, 071702 (2003)
  [hep-ph/0201261].

\bibitem{Balazs:2004ae}
  C.~Balazs, M.~S.~Carena, A.~Menon, D.~E.~Morrissey and C.~E.~M.~Wagner,
  Phys.\ Rev.\  D {\bf 71}, 075002 (2005)
  [hep-ph/0412264].


\bibitem{Lee:2004we}
  C.~Lee, V.~Cirigliano and M.~J.~Ramsey-Musolf,
  Phys.\ Rev.\  D {\bf 71}, 075010 (2005)
  [hep-ph/0412354].
  V.~Cirigliano, S.~Profumo and M.~J.~Ramsey-Musolf,
  JHEP {\bf 0607}, 002 (2006)
  [hep-ph/0603246].

\bibitem{Cirigliano:2006wh}
  V.~Cirigliano, M.~J.~Ramsey-Musolf, S.~Tulin and C.~Lee,
  Phys.\ Rev.\  D {\bf 73}, 115009 (2006)
  [hep-ph/0603058].

\bibitem{Chang:1998uc}
  D.~Chang, W.~Y.~Keung and A.~Pilaftsis,
  Phys.\ Rev.\ Lett.\  {\bf 82}, 900 (1999)
  [Erratum-ibid.\  {\bf 83}, 3972 (1999)]
  [hep-ph/9811202].
  A.~Pilaftsis,
  Phys.\ Lett.\  B {\bf 471}, 174 (1999)
  [hep-ph/9909485].
  J.~R.~Ellis, J.~S.~Lee and A.~Pilaftsis,
  JHEP {\bf 0810}, 049 (2008)
  [0808.1819 [hep-ph]].

\bibitem{Chang:2005ac}
  D.~Chang, W.~F.~Chang and W.~Y.~Keung,
  Phys.\ Rev.\  D {\bf 71}, 076006 (2005)
  [hep-ph/0503055].

\bibitem{Giudice:2005rz}
  G.~F.~Giudice and A.~Romanino,
  Phys.\ Lett.\  B {\bf 634}, 307 (2006)
  [hep-ph/0510197].

\bibitem{Li:2008kz}
  Y.~Li, S.~Profumo and M.~Ramsey-Musolf,
  Phys.\ Rev.\  D {\bf 78}, 075009 (2008)
  [0806.2693 [hep-ph]].

\bibitem{Pilaftsis:2002fe}
  A.~Pilaftsis,
  Nucl.\ Phys.\  B {\bf 644}, 263 (2002)
  [hep-ph/0207277].

\bibitem{Konstandin:2005cd}
  T.~Konstandin, T.~Prokopec, M.~G.~Schmidt and M.~Seco,
  Nucl.\ Phys.\  B {\bf 738}, 1 (2006)
  [hep-ph/0505103].

\bibitem{Li:2008ez}
  Y.~Li, S.~Profumo and M.~Ramsey-Musolf,
  Phys.\ Lett.\  B {\bf 673}, 95 (2009)
  [0811.1987 [hep-ph]].

\bibitem{Carena:1996wj}
  M.~S.~Carena, M.~Quiros and C.~E.~M.~Wagner,
  Phys.\ Lett.\  B {\bf 380}, 81 (1996)
  [hep-ph/9603420].

\bibitem{Delepine:1996vn}
  D.~Delepine, J.~M.~Gerard, R.~Gonzalez Felipe and J.~Weyers,
  Phys.\ Lett.\  B {\bf 386}, 183 (1996)
  [hep-ph/9604440].

\bibitem{Amsler:2008zzb}
  C.~Amsler {\it et al.}  [Particle Data Group],
  Phys.\ Lett.\  B {\bf 667}, 1 (2008).

\bibitem{Carena:2008vj}
  M.~Carena, G.~Nardini, M.~Quiros and C.~E.~M.~Wagner,
  Nucl.\ Phys.\  B {\bf 812}, 243 (2009)
  [0809.3760 [hep-ph]].


\bibitem{Carena:2006gb}
  M.~S.~Carena and A.~Freitas,
  Phys.\ Rev.\  D {\bf 74}, 095004 (2006)
  [hep-ph/0608255].


\bibitem{Demina:1999ty}
  R.~Demina, J.~D.~Lykken, K.~T.~Matchev and A.~Nomerotski,
  Phys.\ Rev.\  D {\bf 62}, 035011 (2000)
  [hep-ph/9910275].

\bibitem{Balazs:2004bu}
  C.~Balazs, M.~S.~Carena and C.~E.~M.~Wagner,
  Phys.\ Rev.\  D {\bf 70}, 015007 (2004)
  [hep-ph/0403224].

\bibitem{Aaltonen:2007sw}
  T.~Aaltonen {\it et al.}  [CDF Collaboration],
  Phys.\ Rev.\  D {\bf 76}, 072010 (2007)
  [0707.2567 [hep-ex]].



\bibitem{Abazov:2008rc}
  V.~M.~Abazov {\it et al.}  [D0 Collaboration],
  Phys.\ Lett.\  B {\bf 665}, 1 (2008)
  [0803.2263 [hep-ex]].


\bibitem{Kraml:2005kb}
  S.~Kraml and A.~R.~Raklev,
  Phys.\ Rev.\  D {\bf 73}, 075002 (2006)
  [hep-ph/0512284];
  S.~Kraml and A.~R.~Raklev,
  AIP Conf.\ Proc.\  {\bf 903}, 225 (2007)
  [hep-ph/0609293].

\bibitem{Bhattacharyya:2008tw}
  N.~Bhattacharyya, A.~Datta and M.~Maity,
  Phys.\ Lett.\  B {\bf 669}, 311 (2008)
  [0807.0994 [hep-ph]].


\bibitem{Martin:2008aw}
  S.~P.~Martin,
  Phys.\ Rev.\  D {\bf 78}, 055019 (2008)
  [0807.2820 [hep-ph]].


\bibitem{Carena:2008mj}
  M.~Carena, A.~Freitas and C.~E.~M.~Wagner,
  JHEP {\bf 0810}, 109 (2008)
  [0808.2298 [hep-ph]].

\bibitem{Hiller:2008wp}
  G.~Hiller and Y.~Nir,
  JHEP {\bf 0803}, 046 (2008)
  [0802.0916 [hep-ph]].

\bibitem{Martin:2008sv}
  S.~P.~Martin,
  Phys.\ Rev.\  D {\bf 77}, 075002 (2008)
  [0801.0237 [hep-ph]];
  S.~P.~Martin and J.~E.~Younkin,
  [0901.4318 [hep-ph]].

\bibitem{Kileng:1995pm}
  B.~Kileng, P.~Osland and P.~N.~Pandita,
  Z.\ Phys.\  C {\bf 71}, 87 (1996)
  [hep-ph/9506455].


\bibitem{Kane:1995ek}
  G.~L.~Kane, G.~D.~Kribs, S.~P.~Martin and J.~D.~Wells,
  Phys.\ Rev.\  D {\bf 53}, 213 (1996)
  [hep-ph/9508265].


\bibitem{Dawson:1996xz}
  S.~Dawson, A.~Djouadi and M.~Spira,
  Phys.\ Rev.\ Lett.\  {\bf 77}, 16 (1996)
  [hep-ph/9603423];
  A.~Djouadi,
  Phys.\ Lett.\  B {\bf 435}, 101 (1998)
  [hep-ph/9806315];
  A.~Djouadi and M.~Spira,
  Phys.\ Rev.\  D {\bf 62}, 014004 (2000)
  [hep-ph/9912476].



\bibitem{Dermisek:2007fi}
  R.~Dermisek and I.~Low,
  Phys.\ Rev.\  D {\bf 77}, 035012 (2008)
  [hep-ph/0701235].

\bibitem{Low:2009nj}
  I.~Low and S.~Shalgar,
  JHEP {\bf 0904}, 091 (2009)
  [0901.0266 [hep-ph]].


\bibitem{Carena:2008rt}
  M.~Carena, G.~Nardini, M.~Quiros and C.~E.~M.~Wagner,
  JHEP {\bf 0810}, 062 (2008)
  [0806.4297 [hep-ph]].

\bibitem{Carena:2002es}
  M.~S.~Carena and H.~E.~Haber,
  Prog.\ Part.\ Nucl.\ Phys.\  {\bf 50}, 63 (2003)
  [hep-ph/0208209];


\bibitem{Djouadi:2005gi}
  A.~Djouadi,
  Phys.\ Rept.\  {\bf 457}, 1 (2008)
  [hep-ph/0503172];
  A.~Djouadi,
  Phys.\ Rept.\  {\bf 459}, 1 (2008)
  [hep-ph/0503173].

\bibitem{Feng:1999mn}
  J.~L.~Feng, K.~T.~Matchev and T.~Moroi,
  Phys.\ Rev.\ Lett.\  {\bf 84}, 2322 (2000)
  [hep-ph/9908309].
  J.~L.~Feng, K.~T.~Matchev and T.~Moroi,
  Phys.\ Rev.\  D {\bf 61}, 075005 (2000)
  [hep-ph/9909334].

\bibitem{Wells:2003tf}
  J.~D.~Wells,
  [hep-ph/0306127];
  J.~D.~Wells,
  Phys.\ Rev.\  D {\bf 71}, 015013 (2005)
  [hep-ph/0411041].


\bibitem{ArkaniHamed:2004fb}
  N.~Arkani-Hamed and S.~Dimopoulos,
  JHEP {\bf 0506}, 073 (2005)
  [hep-th/0405159];
  G.~F.~Giudice and A.~Romanino,
  Nucl.\ Phys.\  B {\bf 699}, 65 (2004)
  [Erratum-ibid.\  B {\bf 706}, 65 (2005)]
  [hep-ph/0406088];
  N.~Arkani-Hamed, S.~Dimopoulos, G.~F.~Giudice and A.~Romanino,
  Nucl.\ Phys.\  B {\bf 709}, 3 (2005)
  [hep-ph/0409232].



\bibitem{Lee:2003nta}
  J.~S.~Lee, A.~Pilaftsis, M.~S.~Carena, S.~Y.~Choi, M.~Drees, J.~R.~Ellis and C.~E.~M.~Wagner,
  Comput.\ Phys.\ Commun.\  {\bf 156}, 283 (2004)
  [hep-ph/0307377];
  J.~R.~Ellis, J.~S.~Lee and A.~Pilaftsis,
  Mod.\ Phys.\ Lett.\  A {\bf 21}, 1405 (2006)
  [hep-ph/0605288];
  J.~S.~Lee, M.~Carena, J.~Ellis, A.~Pilaftsis and C.~E.~M.~Wagner,
  Comput.\ Phys.\ Commun.\  {\bf 180}, 312 (2009)
  [0712.2360 [hep-ph]].

\bibitem{Espinosa:1993yi}
  J.~R.~Espinosa, M.~Quiros and F.~Zwirner,
  Phys.\ Lett.\  B {\bf 307}, 106 (1993)
  [hep-ph/9303317];
  A.~Brignole, J.~R.~Espinosa, M.~Quiros and F.~Zwirner,
  Phys.\ Lett.\  B {\bf 324}, 181 (1994)
  [hep-ph/9312296].


\bibitem{Carena:2002bb}
  M.~S.~Carena, J.~R.~Ellis, S.~Mrenna, A.~Pilaftsis and C.~E.~M.~Wagner,
  Nucl.\ Phys.\  B {\bf 659}, 145 (2003)
  [hep-ph/0211467].

\bibitem{Harlander:2004tp}
  R.~V.~Harlander and M.~Steinhauser,
  JHEP {\bf 0409}, 066 (2004)
  [hep-ph/0409010].

\bibitem{Degrassi:2008zj}
  G.~Degrassi and P.~Slavich,
  Nucl.\ Phys.\  B {\bf 805}, 267 (2008)
  [0806.1495 [hep-ph]].


\bibitem{Abbiendi:2003sc}
  G.~Abbiendi {\it et al.}  [OPAL Collaboration],
  Eur.\ Phys.\ J.\  C {\bf 35}, 1 (2004)
  [hep-ex/0401026].


\bibitem{Ball:2007zza}
  G.~L.~Bayatian {\it et al.}  [CMS Collaboration],
  J.\ Phys.\ G {\bf 34}, 995 (2007).


\bibitem{Aad:2009wy}
  G.~Aad {\it et al.}  [The ATLAS Collaboration],
  [0901.0512 [hep-ex]].


\bibitem{Djouadi:1997xx}
  A.~Djouadi, J.~L.~Kneur and G.~Moultaka,
  Phys.\ Rev.\ Lett.\  {\bf 80}, 1830 (1998)
  [hep-ph/9711244].

\bibitem{Dedes:1999ku}
  A.~Dedes and S.~Moretti,
  Eur.\ Phys.\ J.\  C {\bf 10}, 515 (1999)
  [hep-ph/9904491].
  H.~F.~Heath, C.~Lynch, S.~Moretti and C.~H.~Shepherd-Themistocleous,
  [0901.1676 [hep-ph]].

\bibitem{Noble:2007kk}
  A.~Noble and M.~Perelstein,
  Phys.\ Rev.\  D {\bf 78}, 063518 (2008)
  [0711.3018 [hep-ph]].


\bibitem{cdfreach}
http://www-cdf.fnal.gov/physics/new/hdg/

\bibitem{Aaltonen:2008ec}
  T.~Aaltonen {\it et al.}  [CDF Collaboration],
  [0809.3930 [hep-ex]].

\bibitem{Abazov:2009kq}
  V.~M.~Abazov {\it et al.}  [The D0 Collaboration],
  [0901.1887 [hep-ex]].


\bibitem{Zeppenfeld:2000td}
  D.~Zeppenfeld, R.~Kinnunen, A.~Nikitenko and E.~Richter-Was,
  Phys.\ Rev.\  D {\bf 62}, 013009 (2000)
  [hep-ph/0002036];
  D.~Zeppenfeld,
in {\it Proc. of the APS/DPF/DPB Summer Study on the Future of Particle Physics (Snowmass 2001) } ed. N.~Graf,
{\it In the Proceedings of APS / DPF / DPB Summer Study on the Future of Particle Physics (Snowmass 2001), Snowmass, Colorado, 30 Jun - 21 Jul
2001, pp P123}
  [hep-ph/0203123];

\bibitem{Belyaev:2002ua}
  A.~Belyaev and L.~Reina,
  JHEP {\bf 0208}, 041 (2002)
  [hep-ph/0205270].



\bibitem{Duhrssen:2004cv}
  M.~Duhrssen, S.~Heinemeyer, H.~Logan, D.~Rainwater, G.~Weiglein and D.~Zeppenfeld,
  Phys.\ Rev.\  D {\bf 70}, 113009 (2004)
  [hep-ph/0406323];
  M.~Duhrssen, S.~Heinemeyer, H.~Logan, D.~Rainwater, G.~Weiglein and D.~Zeppenfeld,
  [hep-ph/0407190].

\bibitem{Actis:2008ts}
  S.~Actis, G.~Passarino, C.~Sturm and S.~Uccirati,
  Phys.\ Lett.\  B {\bf 670}, 12 (2008)
  [0809.1301 [hep-ph]];
  S.~Actis, G.~Passarino, C.~Sturm and S.~Uccirati,
  Nucl.\ Phys.\  B {\bf 811}, 182 (2009)
  [0809.3667 [hep-ph]].

\bibitem{Anastasiou:2008tj}
  C.~Anastasiou, R.~Boughezal and F.~Petriello,
  JHEP {\bf 0904}, 003 (2009)
  [0811.3458 [hep-ph]].

\bibitem{deFlorian:2009hc}
  D.~de Florian and M.~Grazzini,
  Phys.\ Lett.\  B {\bf 674}, 291 (2009)
  [0901.2427 [hep-ph]].


\bibitem{Ahrens:2008qu}
  V.~Ahrens, T.~Becher, M.~Neubert and L.~L.~Yang,
  Phys.\ Rev.\  D {\bf 79}, 033013 (2009)
  [0808.3008 [hep-ph]];
  V.~Ahrens, T.~Becher, M.~Neubert and L.~L.~Yang,
  [0809.4283 [hep-ph]].

\bibitem{Djouadi:2003jg}
  A.~Djouadi and S.~Ferrag,
  Phys.\ Lett.\  B {\bf 586}, 345 (2004)
  [hep-ph/0310209].

\bibitem{Stirling:2008sj}
For a recent review, see:\\
  W.~J.~Stirling,
  [0812.2341 [hep-ph]].

\bibitem{Dittmar:2009ii}
  M.~Dittmar {\it et al.},
  [0901.2504 [hep-ph]].


\bibitem{Gianotti:2002xx}
  F.~Gianotti {\it et al.},
  Eur.\ Phys.\ J.\  C {\bf 39}, 293 (2005)
  [hep-ph/0204087].




\bibitem{Kribs:2007nz}
  G.~D.~Kribs, T.~Plehn, M.~Spannowsky and T.~M.~P.~Tait,
  Phys.\ Rev.\  D {\bf 76}, 075016 (2007)
  [0706.3718 [hep-ph]].

\bibitem{Arik:2007vi}
  E.~Arik, S.~A.~Cetin and S.~Sultansoy,
  Balk.\ Phys.\ Lett.\  {\bf 15N4}, 1 (2007)
  [0708.0241 [hep-ph]].

\bibitem{Fok:2008yg}
  R.~Fok and G.~D.~Kribs,
  Phys.\ Rev.\  D {\bf 78}, 075023 (2008)
  [0803.4207 [hep-ph]].


\bibitem{Morrissey:2003sc}
  D.~E.~Morrissey and C.~E.~M.~Wagner,
  Phys.\ Rev.\  D {\bf 69}, 053001 (2004)
  [hep-ph/0308001].

\bibitem{Ham:2008xf}
  S.~W.~Ham, T.~Hur, P.~Ko and S.~K.~Oh,
  J.\ Phys.\ G {\bf 35}, 095007 (2008)
  [0801.2361 [hep-ph]].



\bibitem{Han:2003gf}
  T.~Han, H.~E.~Logan, B.~McElrath and L.~T.~Wang,
  Phys.\ Lett.\  B {\bf 563}, 191 (2003)
  [Erratum-ibid.\  B {\bf 603}, 257 (2004)]
  [hep-ph/0302188];
  T.~Han, H.~E.~Logan and L.~T.~Wang,
  JHEP {\bf 0601}, 099 (2006)
  [hep-ph/0506313].

\bibitem{Belyaev:2005ct}
  A.~Belyaev, A.~Blum, R.~S.~Chivukula and E.~H.~Simmons,
  Phys.\ Rev.\  D {\bf 72}, 055022 (2005)
  [hep-ph/0506086].

\bibitem{Cacciapaglia:2009ky}
  G.~Cacciapaglia, A.~Deandrea and J.~Llodra-Perez,
  JHEP {\bf 0906}, 054 (2009)
  [0901.0927 [hep-ph]].

\bibitem{Barger:2009me}
  V.~Barger, H.~E.~Logan and G.~Shaughnessy,
  Phys.\ Rev.\  D {\bf 79}, 115018 (2009)
  [0902.0170 [hep-ph]].


\bibitem{Bhattacharyya:2009nb}
  G.~Bhattacharyya and T.~S.~Ray,
  [0902.1893 [hep-ph]].



\end{thebibliography}
\end{document}